\documentclass[journal,10pt,twocolume]{IEEEtran}

\usepackage{booktabs}
\usepackage{amsmath}
\usepackage{amsthm}
\usepackage{multirow}
\usepackage{color,cite}
\usepackage{subfigure}
\usepackage{mathrsfs}
\usepackage{amssymb}
\usepackage{bm}
\usepackage{graphicx}
\usepackage{amsfonts}
\usepackage{algorithm}
\usepackage{algorithmic}
\usepackage{epstopdf}
\usepackage{makecell}
\usepackage{geometry}
\usepackage{hyperref}
\hypersetup{hidelinks,colorlinks=true,urlcolor=blue,pdfstartview=Fit,breaklinks=true}
\begin{document}

\bibliographystyle{IEEEtran}

\title{\huge Two-timescale Beamforming Optimization for Downlink Multi-user Holographic MIMO Surfaces}
\author{
    Haochen Wu, Yuanbin Chen, Yang Ming and Zhaocheng Wang, \emph{Fellow, IEEE}
    \thanks{Copyright (c) 2015 IEEE. Personal use of this material is permitted. However, permission to use this material for any other purposes must be obtained from the IEEE by sending a request to pubs-permissions@ieee.org.}
    \thanks{This work was supported in part by National Key R\&D Program of China under Grant 2018YFB1801501. (\emph{Corresponding author: Zhaocheng Wang.})}
    \thanks{H. Wu is with the Department of Electronic Engineering, Tsinghua University, Beijing 100084, China (e-mail: wuhc23@mails.tsinghua.edu.cn).}
    \thanks{Y. Chen is with the State Key Laboratory of Networking and Switching Technology, Beijing University of Posts and Telecommunications, Beijing 100876, China (e-mail: chen\_yuanbin@163.com).}
    \thanks{Y. Ming is with the Shenzhen International Graduate School, Tsinghua University, Shenzhen 518055, China (e-mail: mingy20@mails.tsinghua.edu.cn).}
    \thanks{Z. Wang is with the Department of Electronic Engineering, Tsinghua University, Beijing 100084, China, and also with the Shenzhen International Graduate School, Tsinghua University, Shenzhen 518055, China (e-mail:zcwang@tsinghua.edu.cn).}
    \vspace{-10mm}
} 
\maketitle

\begin{abstract}
    Benefiting from the rapid development of metamaterials and metasurfaces,
    the holographic multiple-input and multiple-output surface (HMIMOS) has been regarded as a promising solution for future wireless networks recently.
    By densely packing numerous radiation elements together, HMIMOS is capable of realizing accurate beamforming with low hardware complexity.
    However, enormous radiation elements on the HMIMOS lead to high computational complexity and signaling overhead when applying traditional beamforming schemes relying on instantaneous channel state information (CSI).
    To tackle this problem, we propose a two-timescale optimization scheme to minimize the required transmission power under the constraint of all users' quality-of-service (QoS).    
    Specifically, the beampatterns at the base station (BS) and the user equippment (UE) are optimized over the slowly changing statistical CSI
    based on the constrained stochastic successive convex approximation (CSSCA) algorithm.
    Then, the instantaneous CSI is utilized to design the precoding matrix 
    in order to ensure the system performance without significant increase in computational cost,
    due to the small number of feeds on the HMIMOS.
    Simulation results demonstrate the effectiveness of our proposed method compared to other baselines.
\end{abstract}


\section{Introduction} \label{S1}
Following the recent breakthroughs on the fabrication of programmable metamaterials,
the holographic multiple-input and multiple-output surface (HMIMOS) paradigm has attracted increasing attention for its flexible and accurate beam steering capabilities, as well as low manufacturing and hardware costs.
To be specific, inlaid with numerous radiation elements whose spacing is smaller than half wavelength of incident electromagnetic (EM) waves,
HMIMOS is able to realize the continuous aperture and form sharp beams with weak sidelobes.
Moreover, based on the holographic principle and advanced programmable metamaterial technology,
HMIMOS can easily reconfigure its holographic pattern and control its EM response with low cost,
hence generating the desired beam directions flexibly \cite{HMIMOS}.

HMIMOS has been investigated in various aspects recently, such as channel modeling \cite{Modeling},  hardware design \cite{Hardware1,Hardware2} and localization application \cite{Localization}.
Particularly, regarding the holographic beamforming techniques, in \cite{Beamforming1}, a hybrid beamforming scheme is proposed to achieve accurate multi-beam steering in a HMIMOS-assisted multi-user system.
In \cite{Beamforming2}, the satellite communication networks are considered, and a holographic beamforming algorithm is proposed to maximize the system sum rate.
Moreover, the discrete amplitude-controlled HMIMOS is investigated in \cite{Beamforming3}, demonstrating a lower bound for the sum rate in quantized HMIMOS-assisted communication systems.
However, obtaining instantaneous channel state information (CSI) is challenging in holographic MIMO scenarios,
since numerous radiation elements on the HMIMOS lead to larger CSI size,
and the signaling overhead which is typically related to the scale of CSI becomes also higher.
In addition, enormous radiation elements lead to high computational complexity during beamforming optimization
when applying traditional beamforming algorithms.
Therefore, the conventional beamforming methods requiring instantaneous CSI like \cite{Beamforming1,Beamforming2} becomes impractical,
which makes practical beamforming in the HMIMOS-assisted system a challenging task.

In this paper, a two-timescale (TTS) beamforming optimization method is proposed in a downlink multi-user HMIMOS-assisted communication system
to reduce the computational complexity and the signaling overhead in comparison to the beamforming schemes relying upon the instantaneous CSI.
We aim to minimize the average transmit power at the BS, subject to the quality-of-service (QoS) constraints of all users on the achievable spectral efficiency.
On one hand, the long-term beampatterns at the BS and the UE side are optimized according to the slow-varying statistical CSI.
On the other hand, the short-term precoding matrix at the BS side is properly designed in different time slots based on all users' reconfigured instantaneous CSI with optimized long-term beampatterns,
therefore reducing the average transmit power of the BS effectively.
Especially, the overhead for instantaneous CSI acquisition utilized in the precoding design is proportional to the number of feeds attached to the HMIMOS,
which is much less than the number of radiation elements.
As a result, the complexity can not be introduced by leveraging instantaneous CSI in the precoding step.
In order to solve the formulated optimization problem spanning over different timescales,
the constrained stochastic successive convex approximation (CSSCA) algorithm is employed.
Simulation results reveal that our proposed method is able to achieve a proper trade-off between the system transmit power and the overhead of channel acquisition.

\section{System Model and Problem Formulation} \label{S2}
\subsection{System Model}
\begin{figure}[tp!]
    \begin{center}
        \includegraphics[width=0.4\textwidth]{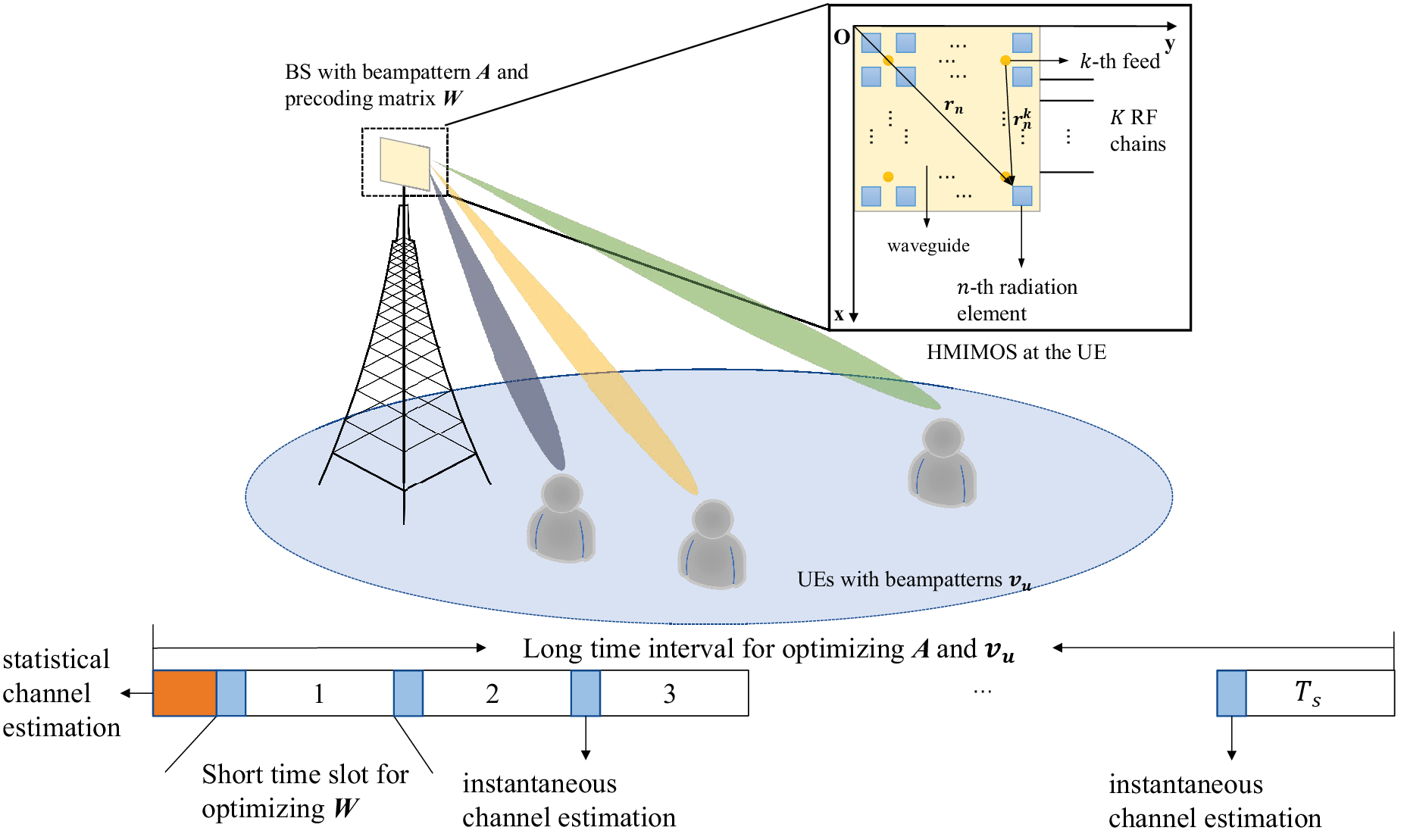}
    \end{center}
    \vspace{-3mm}
    \caption{Diagram of TTS method.}
    \label{Model}
    \vspace{-5mm}
\end{figure}

As shown in Fig.~\ref{Model}, we consider a downlink multi-user communication system between one BS and a group of $U$ UEs denoted by $\mathcal{U}=\{1,2,\cdots ,U\}$.
Each user is served by one data stream.
Moreover, we assume that both BS and UEs are equipped with HMIMOS.
The HMIMOS is a kind of special leaky-wave antenna consisting of feeds, a planar array of amplitude-controlled radiation elements and a parallel-plate waveguide,
which is depicted in Fig.~\ref{Model}.
The feeds are embedded in the bottom layer of the HMIMOS in order to generate reference waves, which propagate on the waveguide and carry the radiating signals.
Meanwhile, the radiation elements are located on the surface of the waveguide.
Excited by the reference wave,
the amplitude-controlled radiation elements can intelligently control their electromagnetic response,
thus generating direct and sharp beams with low sidelobes.

As for the key features of HMIMOS, the number of antenna elements is much more than that in the conventional MIMO,
which leads to better beamforming performance.
In addition, these radiation elements are all amplitude-controlled since the HMIMOS is a kind of leaky-wave antenna,
while the phase control method is usually adopted in the conventional MIMO.
When compared with reconfigurable intelligent surfaces (RISs),
HMIMOS is different in terms of its physical structure and operating mechanism.
Additionally, HMIMOS is generally used as the active transmitting and receiving antenna
in order to reduce the complexity of feeding networks,
while the RIS is widely utilized as passive relays to extend cell coverage.

Specifically, in our considered system, the HMIMOS at the BS consists of $N=N_t^2$ radiation elements and $K$ feeds,
while those at the UEs are comprised of $M=N_r^2$ radiation elements and one feed.
It is required that the number of feeds at the BS should be more than the number of active data streams, i.e. $K>U$,
and the spacing of radiation elements is below half of the wavelength $\lambda$ of the incident EM wave \cite{Beamforming1}.

Since the HMIMOS can not perform digital signal processing, digital beamforming is performed at the baseband and at the BS.
Accordingly, the received signal $y_u$ of the $u$-th UE can be expressed as
\begin{equation} \label{System}
    \begin{aligned}
        y_u & =\boldsymbol{v}_u^H\bm{B}\bm{H}_u^H\bm{A}\bm{\Theta}\boldsymbol{w}_u x_u                                                               \\
            & +\boldsymbol{v}_u^H\bm{B}\bm{H}_u^H\bm{A}\bm{\Theta}\sum_{u'\neq u}\boldsymbol{w}_{u'}x_{u'}+\boldsymbol{v}_u^H\bm{B}\boldsymbol{z}_u,
    \end{aligned}
\end{equation}
where $x_u$ is the intended signal for the $u$-th UE, and $\boldsymbol{w}_u \in \mathbb{C}^{K \times 1}$ is the $u$-th column of the precoding matrix $\bm{W}$ at the BS.

Moreover, we denote $\boldsymbol{\alpha} \in \mathbb{R}^{N \times 1}$ as the beampattern of the HMIMOS at the BS,
and each element in $\boldsymbol{\alpha}$ represents the gain of one corresponding radiation element on the HMIMOS at the transmitter side,
which is obtained based on the holographic principle.
For the $n$-th radiation element on the HMIMOS, we denote its position as $\boldsymbol{r}_{n}$,
and its distance vector from the $k$-th feed as $\boldsymbol{r}_n^k$.
In addition, the propagation vector of the object beam $\Psi_{obj}$ 
and the reference waves generated by all feeds $\Psi_{ref}$ 
are regarded as $\boldsymbol{k}_s$ and $\boldsymbol{k}_f$, respectively.
Therefore, the interference $\Psi_{intf}$ between the object beam and the reference wave,
which is named as the beampattern, can be expressed as
\begin{equation}
    \begin{aligned}
        \Psi_{intf}&=\Psi_{obj}\Psi_{ref}^*=e^{-j(\boldsymbol{k}_f\boldsymbol{r}_n-\boldsymbol{k}_s\boldsymbol{r}_n^k)}.
    \end{aligned}
\end{equation}
If the acquired beampattern is excited by the reference wave $\Psi_{ref}$,
the generated beam satisfies $\Psi_{ref}\Psi_{intf} \propto \Psi_{obj}|\Psi_{ref}|^2$,
whose direction is exactly the same as the object beam.
Note that the radiation elements are all amplitude-controlled,
$\Re(\Psi_{intf})$ is usually chosen to parameterize the radiation amplitude to generate the object beam.
In order to avoid the negative values, $\Re(\Psi_{intf})$ is normalized to $[0,1]$.
Therefore, the amplitude for $\alpha(\boldsymbol{r}_n^k)$ can be expressed by
\begin{equation}
    \alpha(\boldsymbol{r}_n^k)=\frac{\Re(\Psi_{intf}(\boldsymbol{r}_n^k))+1}{2}.
\end{equation}
We denote $\bm{A}=\text{diag}(\boldsymbol{\alpha})=\text{diag}([\alpha_1,\alpha_2, \cdots, \alpha_N])$,
where $\alpha_n=\frac{\sum_{i=1}^K\alpha(\boldsymbol{r}_n^k)}{K}$.
Similarly, for the $u$-th UE, the beampattern of the HMIMOS is $\boldsymbol{v}_u \in \mathbb{R}^{M \times 1}$,
which represents the gain weight of each radiation elements of different HMIMOS at the receiver side,
and $\bm{V}=[\boldsymbol{v}_1,\boldsymbol{v}_2, \cdots, \boldsymbol{v}_U]$.
According to the basic principles of the HMIMOS, the elements in $\boldsymbol{v}_u$ are also real numbers in $[0,1]$ \cite{HMIMOS}.

Furthermore, $\bm{\Theta}$ is a $N \times K$ complex matrix composed of element $\{e^{-j\boldsymbol{k}_s\boldsymbol{r}_n^k}\}$ 
which represents the phase difference between the $n$-th radiation element and the $k$-th feed at the transmitter side.
Both $\bm{A}$ and $\bm{\Theta}$ are dependent on the distance vector $\boldsymbol{r}_n^k$,
which allows the holographic beamforming to be different from the conventional hybrid precoding in massive MIMO systems.
Since the phase matrix is determined once the positions of feeds and radiation elements are fixed,
we assume the phase matrices for all UEs to be the same, which is similarly denoted by $\boldsymbol{\beta} \in \mathbb{C}^{M \times 1}$,
and $\bm{B}=\text{diag}(\boldsymbol{\beta})$.
The additive white Gaussian noise is denoted by $\boldsymbol{z}_u \sim \mathcal{CN}(\boldsymbol{0}, \sigma^2\bm{I}_{U})$.

As for the channel model, the widely used Saleh-Valenzuela model for millimeter-wave communications is adopted \cite{Channel1}.
The channel matrix $\bm{H}_u$ for the $u$-th UE consisting of one line-of-sight (LoS) path and $L$ non-LoS (NLoS) paths can be expressed as
\begin{equation} \label{Channel}
    \begin{aligned}
        \bm{H}_u= & \beta_u^{(0)}\boldsymbol{a}_r(\omega_u^{(0)},\phi_u^{(0)})\boldsymbol{a}_t(\theta_u^{(0)},\psi_u^{(0)})^H +             \\
                  & \sum_{l=1}^L \beta_u^{(l)}\boldsymbol{a}_r(\omega_u^{(l)},\phi_u^{(l)})\boldsymbol{a}_t(\theta_u^{(l)},\psi_u^{(l)})^H,
    \end{aligned}
\end{equation}
where $\beta_u^{(l)}, l=0,1,\cdots L$ are the complex gains of LoS and NLoS paths,
and $(\omega_u^{(l)},\phi_u^{(l)})$ and $(\theta_u^{(l)},\psi_u^{(l)})$ for $0 \leq l \leq L$ represent the physical angles of arrival (AoAs) and angles of departure (AoDs).
In addition, the antenna array response at the $u$-th UE and the BS are $\boldsymbol{a}_r(\omega_u^{(l)},\phi_u^{(l)})$ and $\boldsymbol{a}_t(\theta_u^{(l)},\psi_u^{(l)})$ for both LoS and NLoS paths, respectively \cite{Beamforming1}.
It is assumed that the spacing between the radiation elements of the HMIMOS at the BS and at all UEs along the $x$-axis and $y$-axis are both $\Delta_s$.
After taking the position of the HMIMOS at both the BS and the UEs into consideration, $\boldsymbol{a}_r(\omega_u^{(l)},\phi_u^{(l)})$ and $\boldsymbol{a}_t(\theta_u^{(l)},\psi_u^{(l)})$ can be given by

\begin{equation} \label{Angle1}
    \begin{aligned}
         & \boldsymbol{a}_r(\omega_u^{(l)},\phi_u^{(l)})                                                        \\
         & =[1,\cdots, e^{jk_f[(i-1)\Delta_s\sin\omega_u^{(l)}\cos\phi_u^{(l)}+(j-1)\Delta_s\cos\phi_u^{(l)}]}, \\
         & \cdots,e^{jk_f[(N_r-1)\Delta_s\sin\omega_u^{(l)}\cos\phi_u^{(l)}+(N_r-1)\Delta_s\cos\phi_u^{(l)}]}], \\
    \end{aligned}
\end{equation}

\begin{equation} \label{Angle2}
    \begin{aligned}
         & \boldsymbol{a}_t(\theta_u^{(l)},\psi_u^{(l)})                                                        \\
         & =[1,\cdots, e^{jk_f[(i-1)\Delta_s\sin\theta_u^{(l)}\cos\psi_u^{(l)}+(j-1)\Delta_s\cos\psi_u^{(l)}]}, \\
         & \cdots,e^{jk_f[(N_t-1)\Delta_s\sin\theta_u^{(l)}\cos\psi_u^{(l)}+(N_t-1)\Delta_s\cos\psi_u^{(l)}]}], \\
    \end{aligned}
\end{equation}
where $k_f=\frac{2\pi}{\lambda}$ represents the wave number of the incident EM wave.

Based on Eq.~(\ref{System}), the received signal-to-interference-plus-noise ratio (SINR) of the $u$-th UE can be expressed as
\begin{equation} \label{SINR}
    \text{SINR}_u=\frac{|\boldsymbol{v}_u^H\bm{B}\bm{H}_u^H\bm{A}\bm{\Theta}\boldsymbol{w}_u|^2}{\sum_{u' \neq u}|\boldsymbol{v}_u^H\bm{B}\bm{H}_u^H\bm{A}\bm{\Theta}\boldsymbol{w}_{u'}|+||\boldsymbol{v}_u^H\bm{B}||^2\sigma^2},
\end{equation}
and the spectral efficiency of the $u$-th UE can be therefore written as $\log_2{(1+\text{SINR}_u)}$.

\subsection{Problem Formulation} \label{S3}
It is worth pointing out that in conventional beamforming methods,
the beampattern is mainly designed based on the instantaneous CSI \cite{Beamforming1, Beamforming2, Beamforming3}.
Nevertheless, with the increase of radiation elements on the HMIMOS,
the size of the instantaneous CSI increases accordingly.
As a result, obtaining instantaneous CSI in the HMIMOS-assisted systems entails high computational cost and signaling overhead.
Therefore, we propose the TTS optimization method in the HMIMOS communication system
in order to decompose the problem into two sub-problems with different timescales \cite{Decomposition},
which is also shown in Fig.~\ref{Model}.

Specifically, we focus on a sufficient long time interval composed of $T_s$ short time slots,
where the statistical CSI remains unchanged in every long time interval,
and the instantaneous CSI stays the same in each short time slot.
In the long time interval, the long-term problem is solved based on the statistical CSI to optimize the holographic beampatterns at both BS and UE sides.
The statistical CSI, e.g. the channel mean and associated AoAs/AoDs,
can be acquired in the beginning of each long time interval, as shown in Fig.~1.
Owing to the slow-varying statistical CSI, the long-term problem requires fewer channel acquisitions than its conventional counterparts.
Therefore, the computational complexity and signaling overhead could be reduced dramatically.
Moreover, in one short time slot, the instantaneous CSI is utilized to optimize the precoding matrix $\bm{W}$ at the BS.
Similarly, the instantaneous CSI including the channel gains and the phase parameters are acquired in the beginning of each short time slot through instantaneous channel estimation.
For accurate channel estimation in holographic MIMO scenarios,
perfect knowledge of statistical CSI is utilized in \cite{Acquisition1} in order to propose a Gram-Schmidt decomposition channel estimation algorithm,
while the deep-learning based approach is introduced in \cite{Acquisition2},
which has high NMSE gains and reduces the pilot overhead considerably.
The computational complexity and signaling overhead of the short-term optimization
are typically related to the scale of the instantaneous CSI,
which is in proportion to the number of feeds $K$ rather than the number of radiation elements $N$.
Since $K$ is much smaller than $N$, leveraging instantaneous CSI in the short-term optimization can ensure the overall system performance
while avoiding high signaling overhead.

\begin{figure}[tp!]
    \begin{center}
        \includegraphics[width=0.4\textwidth]{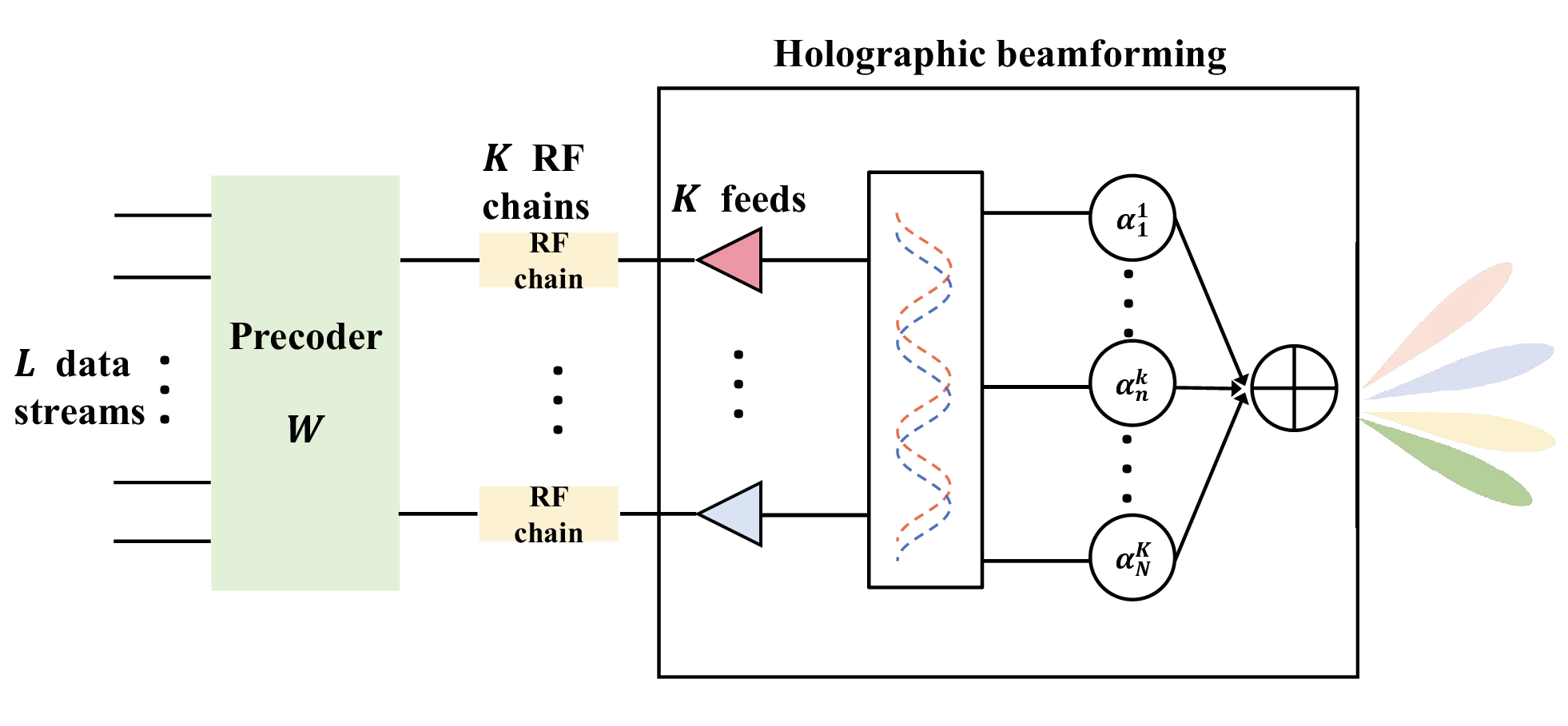}
    \end{center}
    \vspace{-3mm}
    \caption{Data Transmission Process \cite{Holoframe}.}
    \label{Transmission}
    \vspace{-5mm}
\end{figure}

In the considered HMIMOS-assisted system, 
we mainly follow the transmission process in Fig.~\ref{Transmission} \cite{Holoframe}, 
while the distinction is that we optimize the beampattern $\boldsymbol{\alpha},\bm{V}$ and the precoding matrix $\bm{W}$ based on the CSI with different timescales
in order to minimize the expected transmit power with guaranteed QoS for each UE,
which can be written as
\begin{equation} \label{Optimization}
    \begin{aligned}
                     & \min \limits_{\{\boldsymbol{\alpha},\bm{V}\},\bm{W}} \mathbb{E}\{\text{Tr}(\bm{A}\bm{\Theta}\bm{W}\bm{W}^H\bm{\Theta}^H\bm{A}^H)\} \\
        \text{s.t.}~ & \mathbb{E}\{\log_2(1+\text{SINR}_u)\} \geq \delta_u,                                                                               \\
                     & \alpha(i),v_u(i) \in [0,1], \forall u \in \mathcal{U},
    \end{aligned}
\end{equation}
where $\delta_u$ is the threshold of the $u$-th user's spectral efficiency to ensure the QoS for all UEs.

In the long-term problem, the statistical CSI is utilized to optimize the beampattern $\boldsymbol{\alpha}$ and $\bm{V}$ at both the BS and the UE side for given $\widehat{\bm{W}}$.
Accordingly, the long-term problem can be formulated
\begin{equation} \label{Longterm}
    \begin{aligned}
                     & \min \limits_{\boldsymbol{\alpha},\bm{V}} \mathbb{E}\{\text{Tr}(\bm{A}\bm{\Theta}\widehat{\bm{W}}\widehat{\bm{W}}^H\bm{\Theta}^H\bm{A}^H)\} \\
        \text{s.t.}~ & \mathbb{E} \{\log_2(1+\text{SINR}_u)\} \geq \delta_u,                                                                                       \\
                     & \alpha(i),v_u(i) \in [0,1], \forall u \in \mathcal{U}.
    \end{aligned}
\end{equation}

In the short-term problem, the precoding matrix $\bm{W}$ is optimized at each short time slot for given $\widehat{\boldsymbol{\alpha}}$ and $\widehat{\bm{V}}$,
where the instantaneous CSI is leveraged.
The short-term problem can be expressed as
\begin{equation} \label{Shortterm}
    \begin{aligned}
                     & \min \limits_{\bm{W}} \text{Tr}(\widehat{\bm{A}}\bm{\Theta}\bm{W}\bm{W}^H\bm{\Theta}^H\widehat{\bm{A}}^H) \\
        \text{s.t.}~ & \log_2(1+\text{SINR}_u) \leq \delta_u,                                                                    \\
                     & \widehat{\alpha}(i),\widehat{v}_u(i) \in [0,1], \forall u \in \mathcal{U},
    \end{aligned}
\end{equation}
where the expectation operation is removed since the instantaneous CSI is determined in each short time slot.

\section{Methodology Design} \label{S4}
This section details the proposed TTS optimization method,
where the CSSCA algorithm is leveraged to solve the long-term problem with randomly generated channel samples in the long time interval,
while the short-term problem is optimized in each short time slot.

\subsection{Long-term Optimization}
The long-term optimization is performed to obtain the optimal beampattern at the UE and the BS side with fixed precoding matrix at the beginning of every long time interval.
However, the long-term problem (\ref{Longterm}) is non-convex and has an expectation constrained form,
which can not be solved simply.
Consequently, we use the CSSCA algorithm to tackle the difficulty.
The basic idea of CSSCA is to iteratively solving a sequence of convex problems obtained by replacing the objective/constraint function in the original problem with a wide class of surrogate functions \cite{CSSCA}.

To be specific, at each iteration $t$, a set of new channel samples $\{\bm{H}_l^t\}$ is randomly generated from the known statistical CSI.
Since the beampattern matrix $\bm{A}=\text{diag}(\boldsymbol{\alpha})$ at the UE side is a diagonal matrix,
we can firstly utilize the equation $\text{Tr}(\bm{A}\bm{\Theta}\bm{\widehat{\bm{W}}}{\widehat{\bm{W}}}^H\bm{\Theta}^H\bm{A}^H)=\sum_u ||\text{diag}(\bm{\Theta}\widehat{\boldsymbol{w}_u})\boldsymbol{\alpha}||^2$
to convert the problem (\ref{Longterm}) to
\begin{equation} \label{Longterm-sim}
    \begin{aligned}
        \min \limits_{\boldsymbol{\alpha},\bm{V}} f_0(\boldsymbol{\alpha}) & =\mathbb{E}\{g_0(\boldsymbol{\alpha},\widehat{\bm{W}},\bm{H})\}                                           \\
        \text{s.t.}~  f_u(\boldsymbol{\alpha},\bm{V})                      & =\mathbb{E}\{g_u(\boldsymbol{\alpha},\bm{V},\widehat{\bm{W}},\bm{H})\} \leq 0, \forall u \in \mathcal{U},
    \end{aligned}
\end{equation}
where
\begin{subequations}
    \begin{align}
         & g_0(\boldsymbol{\alpha},\widehat{\bm{W}},\bm{H})=\sum_u||\text{diag}(\bm{\Theta}\widehat{\boldsymbol{w}}_u)\boldsymbol{\alpha}||^2, \\
         & g_u(\boldsymbol{\alpha},\bm{V},\widehat{\bm{W}},\bm{H})= \delta_u-\log_2(1+\text{SINR}_u),
    \end{align}
\end{subequations}
and
\begin{equation} \label{SINR-sim}
    \text{SINR}_u=\frac{|\boldsymbol{v}_u^H\bm{B}\bm{H}_u^H\text{diag}(\bm{\Theta}\widehat{\boldsymbol{w}}_u)\boldsymbol{\alpha}|^2}{\sum_{u' \neq u}|\boldsymbol{v}_u^H\bm{B}\bm{H}_u^H \text{diag}(\bm{\Theta}\widehat{\boldsymbol{w}}_{u'})\boldsymbol{\alpha}|^2+||\boldsymbol{v}_u^H\bm{B}||^2\sigma^2}.
\end{equation}

Then, according to the CSSCA algorithm,
the non-convex stochastic functions $f_0(\boldsymbol{\alpha})$ and $f_u(\boldsymbol{\alpha},\bm{V})$ can be approximated with two sequences of surrogate functions for all $1 \leq u \leq U$, i.e.,
\begin{subequations} \label{Surrogate}
    \begin{align}
        \bar{f}_0^t(\boldsymbol{\alpha})        & =f_0^t+(\boldsymbol{f}_{\boldsymbol{\alpha}}^t)^T(\boldsymbol{\alpha}-\boldsymbol{\alpha}^{t-1})+\epsilon_0||\boldsymbol{\alpha}-\boldsymbol{\alpha}^{t-1}||^2,                            \\
        \bar{f}_u^t(\boldsymbol{\alpha},\bm{V}) & =f_u^t+(\boldsymbol{f}_{\boldsymbol{\alpha},u}^t)^T(\boldsymbol{\alpha}-\boldsymbol{\alpha}^{t-1})+(\boldsymbol{f}_{\boldsymbol{v},u}^t)^T(\boldsymbol{v}_u-\boldsymbol{v}_u^{t-1}) \notag \\
                                                & +\epsilon_u(||\boldsymbol{\alpha}-\boldsymbol{\alpha^{t-1}}||^2+||\boldsymbol{v}_u-\boldsymbol{v}_u^{t-1}||^2),
    \end{align}
\end{subequations}
where $\epsilon_0$ and $\epsilon_u$ are constant numbers to guarantee the convexity of the functions.
$f_0^t$ and $f_u^t$ are the average approximation of $f_0(\boldsymbol{\alpha})$ and $f_u(\boldsymbol{\alpha},\bm{V})$ through the channel sample set $\{\bm{H}_l\}$,
which can be updated by
\begin{subequations} \label{Function-approximation}
    \begin{align}
        f_0^t & =(1-\rho^t)f_0^{t-1}+\rho^t\sum_{l=1}^{T_H}\frac{g_0(\boldsymbol{\alpha}^{t-1},\bm{W}_l^t,\bm{H}_l^t)}{T_H},              \\
        f_u^t & =(1-\rho^t)f_u^{t-1}+\rho^t\sum_{l=1}^{T_H}\frac{g_u(\boldsymbol{\alpha}^{t-1},\bm{V}^{t-1},\bm{W}_l^t,\bm{H}_l^t)}{T_H},
    \end{align}
\end{subequations}
where $\rho^t=\frac{1}{(1+t)^{2/3}}$ is a decreasing factor with the increase of the iteration number $t$.
The intermediate variables $\bm{W}_l^t=\bm{W}(\boldsymbol{\alpha}^{t-1};\bm{H_l^{t}})$ are obtained by solving the short-term problem,
while $\boldsymbol{\alpha}^{t-1}$ represents the beampattern at the UE side in the $(t-1)$-th iteration,
and $\bm{H}_l^t$ represents the $l$-th channel sample in the $t$-th iteration.

Moreover, the function vectors $\boldsymbol{f}_{\boldsymbol{\alpha}}^t$, $\boldsymbol{f}_{\boldsymbol{\alpha},u}^t$ and $\boldsymbol{f}_{\boldsymbol{v},u}^t$ are the approximations of the gradients $\nabla_{\boldsymbol{\alpha}}f_0(\boldsymbol{\alpha})$, $\nabla_{\boldsymbol{\alpha}}f_u(\boldsymbol{\alpha},\bm{V})$ and $\nabla_{\boldsymbol{v}_u}f_u(\boldsymbol{\alpha},\bm{V})$ respectively,
which can be given by
\begin{subequations} \label{Gradient-approximation}
    \begin{align}
        \boldsymbol{f}_{\boldsymbol{\alpha}}^t   & =(1-\rho^t)\boldsymbol{f}_{\boldsymbol{\alpha}}^{t-1}+\rho^t\sum_{l=1}^{T_H}\frac{\nabla_{\boldsymbol{\alpha}}g_0(\boldsymbol{\alpha}^{t-1},\bm{W}_l^t,\bm{H}_l^t)}{T_H},                \\
        \boldsymbol{f}_{\boldsymbol{\alpha},u}^t & =(1-\rho^t)\boldsymbol{f}_{\boldsymbol{\alpha},u}^{t-1}+\rho^t\sum_{l=1}^{T_H}\frac{\nabla_{\boldsymbol{\alpha}}g_u(\boldsymbol{\alpha}^{t-1},\bm{V}^{t-1},\bm{W}_l^t,\bm{H}_l^t)}{T_H}, \\
        \boldsymbol{f}_{\boldsymbol{v},u}^t      & =(1-\rho^t)\boldsymbol{f}_{\boldsymbol{v},u}^{t-1}+\rho^t\sum_{l=1}^{T_H}\frac{\nabla_{\boldsymbol{v}_u}g_u(\boldsymbol{\alpha}^{t-1},\bm{V}^{t-1},\bm{W}_l^t,\bm{H}_l^t)}{T_H}.         
    \end{align}
\end{subequations}
The expressions of the gradients in (\ref{Gradient-approximation}) can be easily obtained \cite{Matrix},
which is omitted in this paper.

Based on (\ref{Surrogate})-(\ref{Gradient-approximation}), the long-term problem (\ref{Longterm-sim}) is approximated by solving the following surrogate optimization problem.
\begin{equation} \label{Iteration}
    \begin{aligned}
                     & \min \limits_{\boldsymbol{\alpha},\bm{V}} \bar{f}_0^t(\boldsymbol{\alpha}) \\
        \text{s.t.}~ & \bar{f}_u^t(\boldsymbol{\alpha},\bm{V}) \leq 0, \forall u \in \mathcal{U}.
    \end{aligned}
\end{equation}
As is proven in \cite{CSSCA}, (\ref{Iteration}) is a convex problem.

Finally, with the optimal beampatterns $\bar{\boldsymbol{\alpha}}$ and $\bar{\bm{V}}$ derived from the convex optimization problem (\ref{Iteration}),
$\boldsymbol{\alpha}$ and $\bm{V}$ at the $t$-th iteration are updated as
\begin{subequations} \label{Update}
    \begin{align}
        \boldsymbol{\alpha}^t & =(1-\gamma^t)\boldsymbol{\alpha}^{t-1}+\gamma^t\bar{\boldsymbol{\alpha}}^t, \\
        \bm{V}^t              & =(1-\gamma^t)\bm{V}^{t-1}+\gamma^t\bar{\bm{V}}^t,
    \end{align}
\end{subequations}
where $\gamma^t=\frac{2}{2+t}$ is also a factor which decreases with $t$.

The above iterative long-term optimization algorithm is summarized in Algorithm \ref{Longtermalg}.

\begin{algorithm}[tp!]
    \caption{Long-term CSSCA algorithm}
    \label{Longtermalg}
    \begin{algorithmic}[1]
        \STATE Initialize $\boldsymbol{\alpha}^0$, $\bm{V}^0$ randomly and set $t \leftarrow 1$.
        \STATE \textbf{repeat}
        \STATE Set $\rho^t=\frac{1}{(1+t)^{2/3}}$,$\gamma^t=\frac{2}{2+t}$.
        \STATE Generate a set of $T_H$ channel samples $\{\bm{H}_l^t\}$ according to the statistical CSI.
        \STATE Obtain $\bm{W}_l^t$ based on the channel sample $\bm{H}_l^t$ and $\boldsymbol{\alpha}^{t-1}$ from the previous iteration.
        \STATE Update surrogate functions according to Eq.~(\ref{Surrogate}).
        \STATE Optimize $\bar{\boldsymbol{\alpha}}^t$ and $\bar{\bm{V}}^t$.
        \STATE Update $\boldsymbol{\alpha}^t$ and $\bm{V}^t$ based on Eq.~(\ref{Update}).
        \STATE $t \leftarrow t+1$
        \STATE \textbf{until} the number of iterations $t$ reaches the upper limit of iteration $N_{iter}$.
    \end{algorithmic}
\end{algorithm}

\subsection{Short-term Optimization}
In order to optimize the precoding matrix at the BS,
the short-term problem (\ref{Shortterm}) is performed within every short time slot to guarantee the overall system performance without bringing much computational and signaling overhead.
Through some variable substitutions, the short-term problem can be transformed into a SOCP problem \cite{SOCP}.

$\text{Tr}(\widehat{\bm{A}}\bm{\Theta}\bm{\bm{W}}{\bm{W}}^H\bm{\Theta}^H\widehat{\bm{A}}^H)=\sum_u ||\text{diag}(\bm{\Theta}\boldsymbol{w}_u)\widehat{\boldsymbol{\alpha}}||^2$ is utilized again,
and the short-term problem is therefore converted to
\begin{equation} \label{Shortterm-sim}
    \begin{aligned}
                     & \min \limits_{\bm{W}} \sum_u ||\text{diag}(\bm{\Theta}\boldsymbol{w}_u)\widehat{\boldsymbol{\alpha}}||^2                                                                                                                                                                                 \\
        \text{s.t.}~ & \frac{|\widehat{\boldsymbol{v}}_u^H\bm{B}\bm{H}_u^H\widehat{\bm{A}}\bm{\Theta}\boldsymbol{w}_u|^2}{\sum_{u' \neq u}|\widehat{\boldsymbol{v}}_u^H\bm{B}\bm{H}_u^H\widehat{\bm{A}}\bm{\Theta}\boldsymbol{w}_{u'}|^2+||\widehat{\boldsymbol{v}}_u^H\bm{B}||^2\sigma^2} \geq 2^{\delta_u}-1.
    \end{aligned}
\end{equation}

Let $\bar{\boldsymbol{w}}_u=\text{diag}(\bm{\Theta}\boldsymbol{w}_u)\widehat{\boldsymbol{\alpha}}$,
$\boldsymbol{h_u}=\widehat{\boldsymbol{v}}_u\bm{B}^H\bm{H}_u$, $\eta_u=2^{\delta_u}-1$ and $\bar{\sigma}_u=||\widehat{\boldsymbol{v}}_u^H\bm{B}||\sigma$.
The problem (\ref{Shortterm-sim}) is further simplified to
\begin{equation} \label{Shortterm-sim2}
    \begin{aligned}
                     & \min \limits_{\bm{W}} \sum_u ||\bar{\boldsymbol{w}_u}||^2                                                                                                                       \\
        \text{s.t.}~ & f(u)=\frac{1}{\eta_u\bar{\sigma}^2}|\boldsymbol{h}_u^H\bar{\boldsymbol{w}}_u|^2 \geq \sum_{u' \neq u}\frac{1}{\bar{\sigma}^2}|\boldsymbol{h}_u^H\bar{\boldsymbol{w}}_{u'}|^2+1.
    \end{aligned}
\end{equation}

For the problem (\ref{Shortterm-sim2}), we only need to consider the constraint function,
as the objective function is already convex.
Because the constraint function $f(u)$ only involves the absolute value of variable $\boldsymbol{h}_u^H\bar{\boldsymbol{w}}_u$,
adding a phase factor $e^{j\theta}$ to $\bar{\boldsymbol{w}}_u$ will not cause any difference to the value of $f(u)$.
In this way, $\boldsymbol{h}_u^H\bar{\boldsymbol{w}}_u$ can be considered to be real and positive by adjusting the phase factor added to $\bar{\boldsymbol{w}}_u$.
Next, by taking square roots of both sides of the constraint function, we can express the constraint of the problem (\ref{Shortterm-sim2}) as
\begin{equation}
    \begin{aligned}
         & \frac{1}{\sqrt{\eta_u\bar{\sigma}^2}}\sqrt{|\boldsymbol{h}_u^H\bar{\boldsymbol{w}}_u|^2}=\frac{1}{\sqrt{\eta_u\bar{\sigma}^2}}\boldsymbol{h}_u^H\bar{\boldsymbol{w}}_u \geq \\
         & \sqrt{\sum_{u' \neq u}\frac{1}{\bar{\sigma}^2}|\boldsymbol{h}_u^H\bar{\boldsymbol{w}}_{u'}|^2+1}, \forall u \in \mathcal{U},
    \end{aligned}
\end{equation}
which is a SOCP constraint \cite{Decomposition}.
As a result, the short-term problem is converted to a SOCP optimization problem and can be solved with the CVX toolbox \cite{CVX}.

\section{Simulation Results} \label{S5}
In this section, a multi-user system with $U=4$ users is considered, where the HMIMOS are utilized at both the BS and the UE side.
The UEs are evenly distributed around the BS with a distance within 10 m ,
and the height of the BS and the UEs are set to 25 m and 1.5 m , respectively.
As for the channel model, the UEs are all set to have $L=2$ NLoS paths,
and the complex gains of LoS paths are given by $\beta_u^{(0)}=61.4+20\log_{10}(d_{3D})$,
while the gains of NLoS paths can be denoted by $\beta_u^{(l)} \sim \mathcal{CN}(0,0.1(\beta_u^{(0)})^2)$ according to the settings in \cite{Channel2},
where $d_{3D}$ represents the spatial distance between the BS and UEs.
Other corresponding parameters of the system are listed in Table~\ref{Param}.

\begin{table}[tp!]
    \begin{center}
        \caption{System Parameters.}
        \setlength{\tabcolsep}{2mm}
        \begin{tabular}{cc}
            \toprule[0.8pt]
            Parameters                                       & Values              \\
            \toprule[0.8pt]
            Number of radiation elements at the BS $N$  & 256                 \\
            Number of radiation elements at each UE $M$ & 36                  \\
            Number of feeds at the BS $K$                        & 9                   \\
            Central frequency $f_c$                          & 30 GHz              \\
            Bandwidth $W$                                    & 100 MHz             \\
            Spacing between elements $\Delta_s$              & $\lambda /4=2.5$ mm \\
            Noise power spectrum density                     & -169 dBm/Hz         \\
            \toprule[0.8pt]
        \end{tabular}
        \label{Param}
    \end{center}
    \vspace{-7mm}
\end{table}

The following benchmarks are considered for comparisons.
1) \emph{Alternative optimization (AO) with instantaneous CSI}: $\boldsymbol{\alpha}$, $\bm{V}$ and $\bm{W}$ are alternatively optimized based on the instantaneous CSI for each short time slot \cite{TTS}.
2) \emph{One time slot (OTS) optimization}: $\boldsymbol{\alpha}$, $\bm{V}$ and $\bm{W}$ are optimized based on the instantaneous CSI of the first short time slot slot and remain unchanged during the whole long time interval.
3) \emph{TTS optimization without channel sampling}: $\boldsymbol{\alpha}$, $\bm{V}$ and $\bm{W}$ are optimized based on the TTS algorithm while the channel sample set remains unchanged during the whole long time interval.
4) \emph{Random amplitude}: $\boldsymbol{\alpha}$ is randomly generated, while $\bm{V}$ and $\bm{W}$ are jointly optimized in each short time slot with instantaneous CSI and $\boldsymbol{\alpha}$ \cite{TTS}.
5) \emph{SDMA}: The optimization algorithm in \cite{Holoframe}, where the precoding matrix $\bm{W}$ is obtained through ZF precoding based on instantaneous CSI.

\begin{figure}[tp!]
    \begin{center}
        \includegraphics[width=0.38\textwidth]{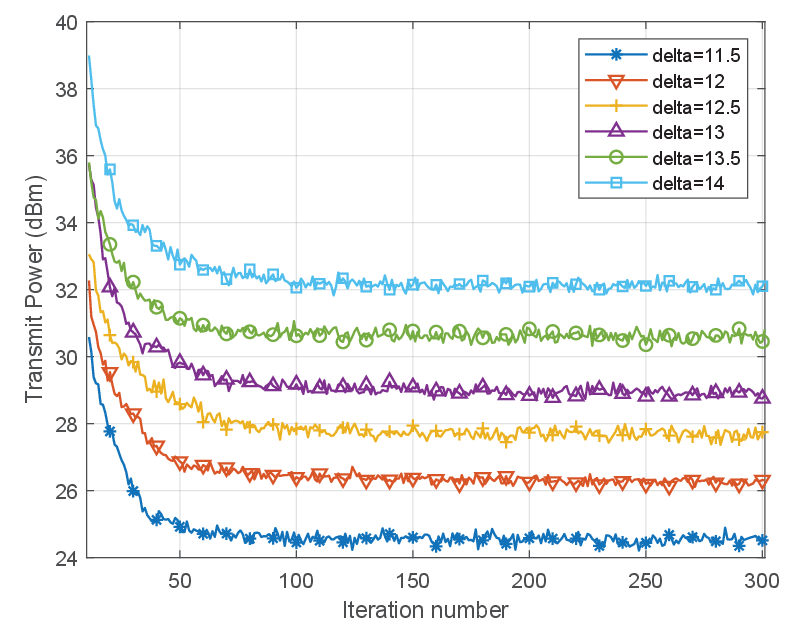}
    \end{center}
    \vspace{-6mm}
    \caption{Convergence behavior of the proposed algorithm under different QoS constraints.}
    \label{Convergence}
    \vspace{-4mm}
\end{figure}

Prior to the performance comparison, we commence by showing the convergence behavior of CSSCA algorithm for the long-term problem as shown in Fig.~\ref{Convergence}.
It can be seen that, our proposed method converges within 300 iterations.
However, due to the non-convexity of the optimization problem and the stochastic nature of the proposed CSSCA algorithm,
the curve of minimum power is not strictly monotonic.
Moreover, with the increase of the user spectral efficiency threshold $\delta_u$, the transmit power in the long-term optimization also converges to a higher value accordingly.

\begin{figure}[tp!]
    \begin{center}
        \includegraphics[width=0.38\textwidth]{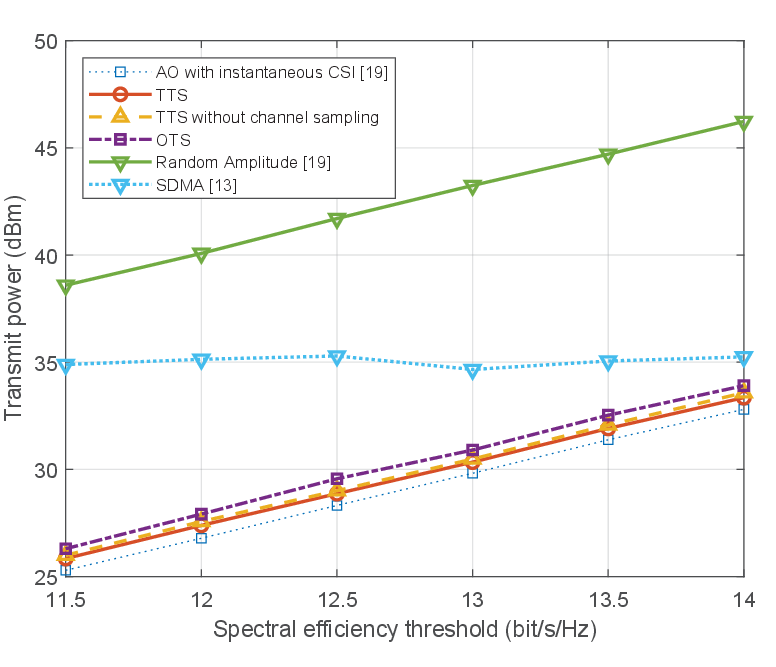}
    \end{center}
    \vspace{-6mm}
    \caption{Minimum transmit power $P$ versus spectral efficiency threshold $\delta_u$.}
    \label{Performance}
    \vspace{-4mm}
\end{figure}

In Fig.~\ref{Performance}, the minimum transmit power achieved by all methods is evaluated over the user spectral efficiency threshold $\delta_u$ with the size of channel sample set $T_H$ fixed to 10, $\epsilon_0$ and $\epsilon_u$ all set to 0.01
and the upper limit of iterations $N_{iter}$ set to 300.
It is clear that our proposed method and all four other benchmarks outperform the \emph{random amplitude} baseline significantly.
Additionally, the transmit power achieved by the proposed TTS algorithm is also smaller than that of the \emph{OTS optimization},
This is because the OTS benchmark only leverages instantaneous CSI in the first short time slot in the whole long time interval,
while our proposed method utilizes both instantaneous CSI and statistical CSI in short-term and long-term optimization, respectively.

When compared to the \emph{TTS optimization without channel sampling} benchmark, our proposed method shows a slight advantage,
which illustrates the importance of the channel sampling step at the beginning of each iteration of CSSCA algorithm.
Additionally, the proposed TTS algorithm also reveals a similar performance in comparison with the \emph{AO with instantaneous CSI} benchmark,
while our method has lower computational complexity and requires less signaling overhead than the benchmark,
since the statistical CSI which is easier to acquire is leveraged to optimize the beampattern instead of the instantaneous CSI.

As for the comparison with the \emph{SDMA} benchmark in \cite{Holoframe},
our proposed method has better performance.
Since the statistical CSI is utilized,
the optimization problem (\ref{Optimization}) in our proposed TTS method has an expectation constrained form
where the instantaneous spectral efficiency of users is not required to be larger than the threshold $\delta_u$ in every short time slot,
allowing for efficient beamforming over different short time slots for the enhanced energy efficiency.
In contrast, the SDMA benchmark is required to firstly satisfy the users' spectral efficiency constraints,
leading to significant performance loss. 

\begin{figure}[tp!]
    \begin{center}
        \includegraphics[width=0.38\textwidth]{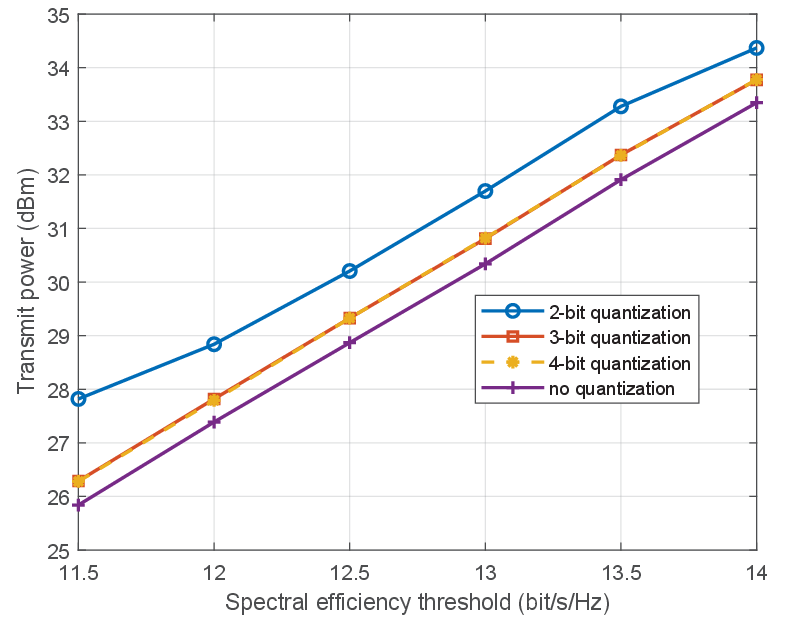}
    \end{center}
    \vspace{-6mm}
    \caption{Performance of TTS algorithm under different quantization bits.}
    \label{Quantization}
    \vspace{-4mm}
\end{figure}

Moreover, due to the hardware resource constraints at both the BS and the UE side,
the performance of our proposed method after $\mu$-law quantization of the beampatterns $\boldsymbol{\alpha}$, $\bm{V}$ and the precoding matrix $\bm{W}$ with different numbers of quantization bits $Q$ is also investigated in Fig.~\ref{Quantization}.
We can conclude from Fig.~\ref{Quantization} that our algorithm has good compability with the quantization operation.
Specifically, when $Q=2$, the quantization will lead to a considerable performance loss to the proposed algorithm.
However, with an increase of $Q$, the performance gap between proposed scheme with or without quantization diminishes to a constant number. and is very close to that with no quantization.

\vspace{-4mm}
\section{Conclusion} \label{S6}
In this paper, we focus on the co-design of the beampattern and the precoding for the downlink multi-user HMIMOS-assisted system.
Since the conventional methods with instantaneous CSI suffers from high computational complexity and signaling overhead,
we conceive a two-stage TTS algorithm, where the statistical CSI is used to optimize the beampatterns instead of the instantaneous CSI.
However, both stages lead to intractable optimization problems,
and thus the CSSCA algorithm is leveraged to solve the formulated non-convex optimization problem.
Simulation results shows that our proposed method achieves a similar performance to the alternative optimization method leveraging the instantaneous CSI,
and is attractive when compared to other baselines.

\vspace{-2mm}
    {

    }
\end{document}